\documentclass[twocolumn,showpacs,prd]{revtex4}
\usepackage{epsfig}
\usepackage{amsmath}
\usepackage{amsfonts}
\usepackage{amssymb}
\usepackage{graphicx}
\usepackage[hyperindex=true,colorlinks=true,pagebackref=false,citecolor=blue,plainpages=false,pdfpagelabels,breaklinks=true,linkcolor=blue,urlcolor=blue,
runcolor=blue,anchorcolor=blue]{hyperref}
\usepackage{microtype}

\def\nn{\nonumber}

\def\J#1#2#3#4{#2 {\bf #3,} #4 (#1)}

\def\PRD{{Phys. Rev.} D}

\def\GRG{Gen. Relat. Grav.}
\def\JMP{J. Math. Phys.}
\def\CQG{Class. Quantum Grav.}
\def\MNRAS{ Mon. Not. R. Astron. Soc.}
\def\APJ{ApJ}
\def\I{{\rm i}}

\begin{document}

\title{{\bf Innermost stable circular orbits around magnetized rotating massive stars}}

\author{Jos\'e D. Sanabria-G\'omez}
\email{jsanabri@uis.edu.co} \affiliation{Escuela de F\'isica,
Universidad Industrial de Santander, A.A. 678, Bucaramanga,
Colombia}

\author{Jos\'e L. Hern\'andez--Pastora}
\email{jlhp@usal.es} \affiliation{Departamento de Matem\'atica Aplicada E. T. S. Ingenier\'ia Industrial de B\'ejar, \\ Universidad de Salamanca, Salamanca, Espa\~na}

\author{F. L. Dubeibe}
\email{fldubeibem@unal.edu.co} \affiliation{Departamento de F\'isica, Universidad Nacional de Colombia, Bogot\'a, Colombia}

\date{\today}

\begin{abstract}
In 1998, Shibata and Sasaki [Phys.~Rev.~D~{\bf 58},~104011 (1998)] presented an approximate analytical formula for the radius of the innermost stable circular orbit (ISCO) of a neutral test particle around a massive, rotating and deformed source. In the present paper, we generalize their expression by including the magnetic dipole moment. We show that our approximate analytical formulas are accurate enough by comparing them with the six-parametric exact solution calculated by Pach\'on {\it et. al.}  [Phys.~Rev.~D~{\bf 73},~104038 (2006)]  along with the numerical data presented by Berti and Stergioulas [MNRAS~{\bf 350},~1416 (2004)] for realistic neutron stars. As a main result, we find that in general, the radius at ISCO exhibits a decreasing behavior with increasing magnetic field. However, for magnetic fields below  $100 {\rm GT}$ the variation of the radius at ISCO is negligible and hence the non-magnetized approximate expression can be used. 
In addition, we derive approximate analytical formulas for angular velocity, energy and angular momentum of the test particle at ISCO.
\end{abstract}


\maketitle


\section{Introduction} \label{Introduction}

The discovery of quasi-periodic oscillations (QPOs) with frequencies around 1 kHz from several low--mass X--ray
binaries (LMXBs)  \cite{kils} has been increasing the interest in the detailed theory of disk accretion
onto neutron stars. Several authors have suggested that at least some of the kHz QPOs may be related to the Kepler frequency at the innermost stable circular orbit (ISCO) of the accretion disk around a neutron star (see {\it e.g.} \cite{nkils,kils}).  Stergioulas {\it et al.} \cite{6} have suggested that the frequency of the co--rotating orbit at ISCO in compact stellar remnant  could be determined through X--ray observations of low--mass X--ray binaries and it could be used to constraint the equation of state (EOS) of ultradense matter. Morsink and Stella \cite{M&S} have remarked the central r\^ole of ISCO in relativistic precession of orbits around neutron stars and Bulik  {\it et al.} \cite{bulik}  have shown that observations are consistent with the assumption that the maximum--frequency QPOs occurs at the ISCO.  The last statement could be used to test general relativity (GR) in the strong--field regime around accreting neutron stars, or even to measure the stellar mass by directly comparing the highest frequency manifest in the X--ray flux with the relativistic formula for the orbital frequency in the ISCO orbit \cite{7}. 

On the other hand, the study of the structure and dynamics of neutron stars endowed with magnetic field in GR is an active, interesting and challenging theoretical issue. The influence of magnetic field on the properties of neutron star rotating at the Kepler frequency has been shown in Ref. \cite{11}. In Ref. \cite{12}, Broderick {\it et al.} have studied the implications of very strong magnetic fields on the structure of neutron stars; in particular, Cardall {\it et. al.} \cite{13} have indicated how magnetic field affects the maximum mass of stars.  In Ref. \cite{101}, the ellipticity of the deformed star due to the rotation and to the magnetic field is calculated, and these two effects are compared to each other within  GR. In addition, the  formulation of deformation of relativistic stars due to the magnetic stress, considering the magnetic fields as perturbations from spherical stars, has also been studied in \cite{10} by means of an analytical treatment assuming weak magnetic fields compared to gravity. The quadrupole deformation of magnetized Newtonian stars was discussed by Chandrasekhar and Fermi \cite{15} and Ferraro \cite{16}. The GR approach was done fully numerical by Bonazzola and Gourgoulhon \cite{18} and Bocquet {\it et al.} \cite{bocquet}, who pointed out that deformations of the stars induced by magnetic fields become appreciable only for fields greater than $10 {\rm GT}$. 

More than one decade ago, Shibata and Sasaki \cite{S&S} (hereafter S\&S) computed an approximate analytical formula for the radius at ISCO on massive rotating  and arbitrarily deformed sources within GR. They considered the r\^ole of the quadrupole moment of mass in physics related to neutron stars (this fact has also been noted by other authors, see for instance \cite{L&P} and references therein), by including multipolar moments of mass up to the $2^4$--pole order in their calculations. Yet, as discussed above, there exist an strong influence of the magnetic field on the structure of the neutron stars and hence is desirable to include also the magnetic field in an approximation as the given by S\&S. In the present paper we tackle this point by following the procedure by S\&S to compute approximate formulas for the ISCO including the magnetic dipole moment. Thereby, our goal is to calculate approximate formulas for the radius, angular velocity, energy and angular momentum at ISCO for massive rotating and deformed sources endowed with magnetic dipole. Due to the importance that these parameters have for magnetized neutron stars (see {\it e.g.} \cite{bocquet, LRR}), we should take into account in our treatment at least the physical parameters of mass, angular momentum, quadrupolar moment of mass, current octupole moment and magnetic dipole. 

The plan of this paper is as follows: In Section \ref{formalism} the general formalism to calculate the ISCO for a neutral particle orbiting around a massive source in GR is presented.  The procedure to compute the approximate formulas for radius , angular velocity, energy and angular momentum at ISCO of a neutral test particle is shown in Section \ref{approx}.  The results along with their analysis are presented in Section \ref{results}. Finally, we present the conclusions of our study.


\section{ISCO AND THE MULTIPOLAR STRUCTURE}\label{formalism}

The  metric describing the geometry of space--time around a stationary and axisymmetric
source, can be written as \cite{P53}
\begin{equation}\label{SASmetric}
ds^2=-f(dt-\omega d\phi)^2 + f^{-1}[e^{2\gamma}(d\rho^2+dz^2)+\rho^2 d\phi^2],
\end{equation}
where $f$, $\gamma$ and $\omega$ are functions of the quasi-cylindrical Weyl-Lewis-Papapetrou coordinates $(t,\rho,z,\phi)$.
In this paper we use geometrized units $c=G=1$. Hence all the physical quantities are measured in units of length [L].

In a standard way, we use the line element (\ref{SASmetric}) to find the geodesic equations for a neutral test particle on the equatorial plane, which reads as follows:
\begin{equation}
\frac{dt}{ds}=\frac{E g_{\varphi\varphi}+Lg_{t\varphi}}{\rho^2}, \qquad \frac{d\varphi}{ds}=-\frac{E g_{t\varphi}+L g_{tt}}{\rho^2},
\end{equation}
\begin{eqnarray}
g_{\varphi\varphi}\left( \frac{d\rho}{dt}\right)^2=&& -(1-E^2\frac{g_{\varphi\varphi}}{\rho^2}-2EL\frac{g_{t\varphi}}{\rho^2}-L^2 \frac{g_{tt}}{\rho^2}) \nonumber \\
=&&-V_{eff}(\rho),
\end{eqnarray}
with $g_{tt=}=-f$, $g_{t\varphi}=f\omega$ and $g_{\varphi\varphi}=-f \omega^2 +\rho^2/f,$ and $V_{eff}(\rho)$ denotes the effective potential.

Circular prograde (or co--rotating) orbits will occur at radius $\rho$ when $V_{eff}=0$ and $dV_{eff}/d\rho=0$, which imposes the following conditions for the angular velocity $\Omega$, the energy $E$ and the angular momentum $L$ of the test particle,
\begin{eqnarray}
\Omega=\frac{-g_{t\varphi,\rho}+\sqrt{(g_{t\varphi,\rho})^2-g_{tt,\rho}g_{\varphi\varphi,\rho}}}{g_{\varphi\varphi,\rho}},\\
E=-\frac{g_{tt}+g_{t\varphi}\Omega}{\sqrt{-g_{tt}-2g_{t\varphi}\Omega-g_{\varphi\varphi}\Omega^2}}, \\
L=\frac{g_{t\varphi}+g_{\varphi\varphi}\Omega}{\sqrt{-g_{tt}-2g_{t\varphi}\Omega-g_{\varphi\varphi}\Omega^2}}.
\end{eqnarray}
The stability of the circular orbit is determined by the sign of
\begin{equation}
\frac{d^2V_{eff}}{d\rho^2}=\frac{1}{\rho^2}\left(  2-E^2 \frac{dg_{\varphi\varphi}}{d\rho^2} -2EL\frac{dg_{t\varphi}}{d\rho^2} -L^2 \frac{dg_{tt}}{d\rho^2} \right), \label{ISCO}
\end{equation}
hereby, ISCOs will occur if and only if $d^2V_{eff}/d\rho^2=0$. It is worth mentioning that the above formulae do not depend on the metric function $\gamma$ and therefore we will left aside this metric function in the rest of the paper.

In order to calculate the metric functions in the electrovacuum case,
we use  the Ernst formulation \cite{ernst}. Via Ernst's procedure,
the Einstein-Maxwell equations can be reformulated in terms of the complex potentials ${\cal E}$ and $\Phi$  as
\begin{eqnarray}
({\rm Re}({\cal E})+ |\Phi|^2) \nabla^2{\cal E}&=& (\nabla {\cal E} + 2 \Phi^*\nabla \Phi )\cdot \nabla {\cal E}, \nonumber \\
({\rm Re}({\cal E})+ |\Phi|^2) \nabla^2 \Phi    &=& (\nabla {\cal E} + 2 \Phi^*\nabla \Phi )\cdot \nabla \Phi. \label{ernst}
\end{eqnarray}
Once the potentials are known, the metric functions $f$ and $\omega$ can be constructed by using
\begin{eqnarray}
{\cal E}=&& f - |\Phi|^2 + \I \, {\rm Im}({\cal E}), \label{f}\\
\omega =&& \int_\rho^\infty d\rho'\frac{\rho'}{f^2} \left[\frac{\partial {\rm Im}({\cal E})}{\partial z } + 2
{\rm Re}(\Phi)  \frac{\partial {\rm Im} (\Phi) }{\partial z } \right. \nonumber \\
&& \left.  - 2 {\rm Im}(\Phi)  \frac{\partial {\rm Re} (\Phi) }{\partial z }\right]_{z={\rm const}}. \label{omega}
\end{eqnarray}

For getting a more intuitive and physical approach, is helpful to change the potentials ${\cal E}$ and $\Phi$ to the potentials $\xi$ and $q$ throughout the following definitions
\begin{equation} {\cal E}:=\frac{1-\xi}{1+\xi}, \qquad \Phi:=\frac{q}{1+\xi}. \label{epsyphi} \end{equation}
This change elucidates the procedure because the potentials $\xi$ and $q$  are related to the gravitational and electromagnetic moment of the source in a very direct way. In order to calculate the multipolar moments of an asymptotically flat space-time, according to the Geroch-Hansen definition \cite{ger70,han74} we use the procedure of Fodor {\it et al.} \cite{F&H&P} with the corrections given by Sotoriou and Apostolatos \cite{S&A}. We need to map the initial 3-metric to a conformal one
$h_{ij} \rightarrow {\tilde h}_{ij} = \Omega^{2}h_{ij}$.
The conformal factor $\Omega$ should satisfy the following conditions:
$\Omega|_{\Lambda} = \tilde{D_{i}}\Omega|_{\Lambda} = 0$ and
$\tilde{D_{i}}\tilde{D_{j}}\Omega|_{\Lambda}= 2h_{ij}|_{\Lambda}$, where $\Lambda$ is the point added to the
initial manifold that represents infinity. $\Omega$ transforms the complex gravitational and electromagnetic
potentials $\xi$ and $q$ into $\tilde{\xi} = \Omega^{-1/2} \xi$ and $\tilde{q} = \Omega^{-1/2} q$
respectively. The conformal factor is given by $\Omega = \bar{r}^{2} = \bar{\rho}^{2} + \bar{z}^{2}$, and the
transformation between unbarred and barred variables reads as
\begin{eqnarray}
\bar{\rho}=\frac{\rho}{\rho^{2}+z^{2}},\quad \bar{z}=\frac{z}{\rho^{2}+z^{2}}, \quad \bar{\phi}=\phi,
\end{eqnarray}
which brings infinity at the origin of the axes $(\bar{\rho},\bar{z}) = (0, 0)$. The potentials $\tilde{\xi}$ and
$\tilde{q}$ can be written in a power series expansion of  $\bar{\rho}$ and $\bar{z}$ as
\begin{eqnarray}\label{xi-q}
\tilde{\xi} = \sum_{i,j=0}^{\infty} a_{ij}\bar{\rho}^{i}\bar{z}^{j},\quad
\tilde{q} = \sum_{i,j=0}^{\infty}  b_{ij}\bar{\rho}^{i}\bar{z}^{j}.
\end{eqnarray}
Due to the analyticity of the potentials at the axis of symmetry, $a_{ij}$ and $b_{ij}$ must vanish when $i$ is odd.
The coefficients in the above power series can be calculated by using the relations \cite{S&A}
\begin{widetext}
\begin{eqnarray}\label{a's}
(r + 2)^{2} a_{r+2,s} &=& -(s + 2)(s + 1)a_{r,s+2} + \sum_{k,l,m,n,p,g}(a_{kl}a^{*}_{mn} - b_{kl}b^{*}_{mn})[a_{pg}(p^{2} + g^{2} - 4p - 5g - 2pk - 2gl - 2)\nn\\
&+& a_{p+2,g-2}(p + 2)(p + 2 - 2k) + a_{p-2,g+2}(g + 2)(g + 1 - 2l)]
\end{eqnarray}
and
\begin{eqnarray}\label{b's}
(r + 2)^{2} b_{r+2,s} &=& -(s + 2)(s + 1)b_{r,s+2} + \sum_{k,l,m,n,p,g} (a_{kl}a^{*}_{mn} - b_{kl}b^{*}_{mn})[b_{pg}(p^{2} + g^{2} - 4p - 5g - 2pk - 2gl - 2)\nn\\
&+& b_{p+2,g-2}(p + 2)(p + 2 - 2k) + b_{p-2,g+2}(g + 2)(g + 1 - 2l)],
\end{eqnarray}
\end{widetext}

where $m = r - k - p, 0 \leq k \leq r, 0 \leq p \leq r - k$ with $k$ and $p$ even, and $n = s - l - g$,
$0 \leq l \leq s + 1$, with $-1 \leq g \leq s - l$. These recurrence relations could build the whole power series
of $\tilde{\xi}$ and $\tilde{q}$ from their values on the axis of symmetry
\begin{eqnarray}
\tilde{\xi}(\bar{\rho}=0,\bar z) = \sum_{i=0}^{\infty} m_{i}\bar{z}^{i},\quad
\tilde{q}(\bar{\rho}=0,\bar z) = \sum_{i=0}^{\infty} q_{i}\bar{z}^{i}\label{q0},
\end{eqnarray}
where  the coefficients in the above series expansion are related to the values of the multipole moments of the space-time
$q_{i}\equiv b_{0i}$ and $m_{i}\equiv a_{0i}$ \cite{F&H&P, S&A}.

\section{APPROXIMATE FORMULAS AT ISCO}
\label{approx}

Following the scheme given by S\&S \cite{S&S} we assume that the only non vanishing multipole moments are the mass $M_0=M$, the angular momentum $M_1= q M^2$, the mass quadrupole $M_2= -{\cal Q}_2 M^3$, the current octupole moment $M_3=- {\cal Q}_3 M^4$, the $2^4$-pole $M_4= {\cal Q}_4 M^5$ and additionally, the magnetic dipole moment ${\cal M}= \mu M^2$, where $q, {\cal Q}_2, {\cal Q}_3, {\cal Q}_4$ and $\mu$ are  dimensionless parameters. In order to keep the approximation consistent up to $\mathcal{O}(\epsilon^4)$, we formally set $q \rightarrow \epsilon q, \quad {\cal Q}_2 \rightarrow \epsilon^2 {\cal Q}_2, \quad {\cal Q}_3 \rightarrow  \epsilon^3 {\cal Q}_3, \quad {\cal Q}_4 \rightarrow \epsilon^4 {\cal Q}_4$ and $\mu \rightarrow \epsilon^2 \mu$.

With the aim to calculate the approximate potentials (\ref{xi-q}) as a truncated power series, we carry out the following steps: (i) We compute the gravitational and electromagnetic multipoles up to order twelve (using the corrected formulas given by Sotiriou and Apostolatos \cite{S&A}), as a function of the coefficients on the symmetry axis, $a_{0,j}$ and $b_{0,j}$.  (ii) By inverting these expressions, we then get the coefficients $a_{0,j}$ and $b_{0,j}$ as a function of the multipoles (see the Appendix and the note \cite{note} at the end of the paper). (iii) Then, we use the expressions for $a_{0,j}$ and $b_{0,j}$ along with the recurrence relations (\ref{a's})-(\ref{b's}), in order to calculate the coefficients $a_{i,j}$ and $b_{i,j}$ up to $\mathcal{O}(\epsilon^4)$.  We do not present here these quantities because of their cumbersome form, but they are available under request to the authors.

Once we know the approximate expressions for the complex potentials  (\ref{xi-q}), it is possible to obtain the approximate expressions of the Ernst potentials ${\cal E}$ and $\Phi$ by applying Eq. (\ref{epsyphi}). Consequently, we compute the metric functions $f$ and $\omega$ (\ref{f})-(\ref{omega}), by expanding in power series of the inverse of $\rho$:
\begin{equation}\label{appf}
f = 1 + \sum_{k=1}^{11} \left( \frac{C_{f,k}}{\rho}\right)^k + \mathcal{O}(\rho^{-12}),
\end{equation}
\begin{equation}\label{appw}
\omega = \sum_{k=1}^{11} \left( \frac{C_{\omega,k}}{\rho}\right)^k + \mathcal{O}(\rho^{-12}),
\end{equation}
where $C_{f,k}$ and $C_{\omega,k}$ are functions of the multipoles.

By using the Eqs. (\ref{appf}) and (\ref{appw})  we can cast the Eq. (\ref{ISCO}) as
\begin{equation}\label{As}
\sum_{k=0}^4 \epsilon^k A_k(\rho,M,q,{\cal Q}_2,{\cal Q}_3,{\cal Q}_4,\mu)=0\,.
\end{equation}
Solving it for $\rho$, we obtain for the circumferencial radius $R=\sqrt{g_{\varphi\varphi}}$ at ISCO:

\begin{eqnarray}
\frac{R_{\rm ISCO}}{6M} =&&   1-0.54433 q -0.22651 q^2 +0.17992 {\cal Q}_2 \nonumber \\
&&-0.00323 \mu^2 -0.23122 q^3 +0.26353 q {\cal Q}_2 \nonumber \\
&&- 0.05318 {\cal Q}_3 - 0.00765 q \mu^2-0.29981 q^4\nonumber \\
&&+0.44887 q^2 {\cal Q}_2 -0.06260 {\cal Q}_2^2 - 0.11325 q {\cal Q}_3 \nonumber \\
&&+0.01546 {\cal Q}_4 - 0.01572 q^2 \mu^2+0.00312 {\cal Q}_2 \mu^2\nonumber \\
&&-0.00004 \mu^4, \label{RISCO}
\end{eqnarray}

and for the angular velocity, the energy and the angular momentum at ISCO:

\begin{eqnarray}
\Omega_{\rm ISCO}= &&\frac{1}{6 \sqrt{6M}} ( 1+0.74860 q + 0.78106 q^2 - 0.23432 {\cal Q}_2 \nonumber \\
&&+ 0.00446 \mu^2 +0.98328 q^3 -0.64492 q {\cal Q}_2  \nonumber \\
&&+ 0.07433 {\cal Q}_3 + 0.01608 q \mu^2 + 1.38828 q^4  \nonumber \\
&&+ 0.12813 {\cal Q}_2^2 + 0.25050 q {\cal Q}_3 - 0.02132 {\cal Q}_4   \nonumber \\
&&- 0.00596 {\cal Q}_2 \mu^2-1.42351 q^2 {\cal Q}_2 + 0.04191 q^2 \mu^2 \nonumber \\
&&+ 0.00007 \mu^4), \label{OMEGA}
\end{eqnarray}
\begin{eqnarray}
E_{\rm ISCO} = &&0.94280 - 0.03208 q - 0.02977 q^2 + 0.00794 {\cal Q}_2 \nonumber \\
&&- 0.00010 \mu^2 - 0.03417 q^3 + 0.01980 q {\cal Q}_2 \nonumber \\
&&- 0.00200 {\cal Q}_3 -0.00035 q \mu^2 -0.04427 q^4 \nonumber \\
&&-0.00331{\cal Q}_2^2 -0.00621 q {\cal Q}_3 + 0.00049 {\cal Q}_4 \nonumber \\
&&+0.00012 {\cal Q}_2 \mu^2 +0.04044 q^2 {\cal Q}_2 - 0.00088 q^2 \mu^2 \nonumber \\
&&+0.00001 \mu^4, \label{E}
\end{eqnarray}
\begin{eqnarray}
\frac{L_{\rm ISCO}}{M} =  &&3.46410 - 0.94281 q - 0.44452 q^2 + 0.18793 {\cal Q}_2 \nonumber \\
&&- 0.00195 \mu^2 -0.39579 q^3 + 0.29982 q {\cal Q}_2 \nonumber \\
&&- 0.03926 {\cal Q}_3 - 0.00519 q \mu^2 - 0.44854 q^4 \nonumber \\
&&-0.05055 {\cal Q}_2^2 - 0.09282 q {\cal Q}_3 + 0.00935 {\cal Q}_4 \nonumber \\
&&+0.00170 {\cal Q}_2 \mu^2 + 0.49506 q^2 {\cal Q}_2 - 0.01099 q^2 \mu^2\nonumber \\
&& - 0.00002 \mu^4.  \label{L}
\end{eqnarray}
\section{RESULTS AND ANALYSIS}\label{results}
First, we verify our results in the vacuum case. We check our formula  for the radius at ISCO (\ref{RISCO}), by setting ${\cal Q}_2=q^2, {\cal Q}_3=q^3, {\cal Q}_4=q^4$ and $\mu=0$, and comparing it with the exact expression given by Bardeen {\it et al.} \cite{bardeen}. Here, we find that the error for  $q<0.5$ is smaller than $1\%$. Next, we compare our expression (\ref{RISCO}) with the six--parametric exact solution given by Pach\'on {\it et al.} \cite{pachon} along with the numerical data given by Berti and Stergioulas  (hereafter B\&S) for selected EOS \cite{B&S}. In this case, for $q<0.3$, the error still being smaller than $0.6\%$ in all the cases. It should be noted at this stage that despite the mistake in Eqs. (2.24), (2.25), (B1) and (B2) of S\&S (see note \cite{note}), the difference with our expressions (\ref{RISCO})--(\ref{L}) is smaller than $0.1\%$ and therefore negligible.

Now, we turn our attention to the electrovacuum case. Starting from the following limiting statements:
(i) Magnetars lose rotational speed very quickly due to their high magnetic field.
(ii) Given their rarity, the possibility to observe a new-born rapidly rotating magnetar is negligible.
(iii) The amount of data for observed magnetars is minimum ({\it cf.} \cite{catalog}) without any data of their higher--order multipolar structure.
(iv) Theoretical studies reporting numerical data of the multipolar structure of magnetars  ({\it cf.} \cite{bocquet}), did not present numerical data for the radius at ISCO nor numerical data for higher mass--rotation multipole moments.
(v) The Pach\'on {\it et al.} solution \cite{pachon} fits very well with realistic numerical interior solutions for slowly rotating neutron stars, possessing an arbitrary magnetic dipole parameter.
Let us to assume that the Pach\'on {\it et al.} solution is a good model for realistic slowly rotating magnetars.

With the aim of testing our approximate formulae, we use the parameters calculated by B\&S for neutron stars, extrapolated to the case of a non-vanishing magnetic dipole. In order to observe the effect of the magnetic field on the radius, hereafter we restrict ourselves to the use of the approximate formula (\ref{RISCO}) in the presence of magnetic dipole $\mu\neq0$, where this approach will be labeled as ``Mag" or in its absence $\mu=0$, where this approach will be labeled as ``Non-Mag".

A rough estimate of the numerical solutions of the Einstein--Maxwell equations presented by Bocquet {\it et al.} (see Table 2 in \cite{bocquet}) for models of rapidly rotating magnetized neutron stars, suggest that the magnetic dipole ${\cal M}$ belongs, in average, to  the interval $0$ to $10^{32}{\rm{A m}^2}$, corresponding to magnetic fields in the range $0$ to $10^{12}{\rm{T}}$.  Hence, from the proportionality between ${\cal M}$ and the magnetic field $B$, the observed magnetic field for magnetars {\it ca.} $10^{11}{\rm{T}}$ \cite{catalog}, should roughly correspond to a magnetic dipole moment of about $10^{31}{\rm{A m}^2}$. Moreover, in natural units the magnetic dipole moment has an order of
\begin{equation}
{\cal M}=\frac{10^{-6}\sqrt{\mu_0 G}}{c^2}{\cal M}_{\rm{S.I}},
\end{equation}
which represents a value of the magnetic dipole of ${\cal M} \sim 1{\rm{km}^2} $ for observed magnetars (see for instance the paragraph around Eq. (20) of Ref. \cite{pachon}).

\begin{figure}[h!]
\begin{center}
\begin{tabular}{cc}
\hspace{0.7cm}(a)&(b)\\
{\includegraphics[angle=270,scale=0.65]{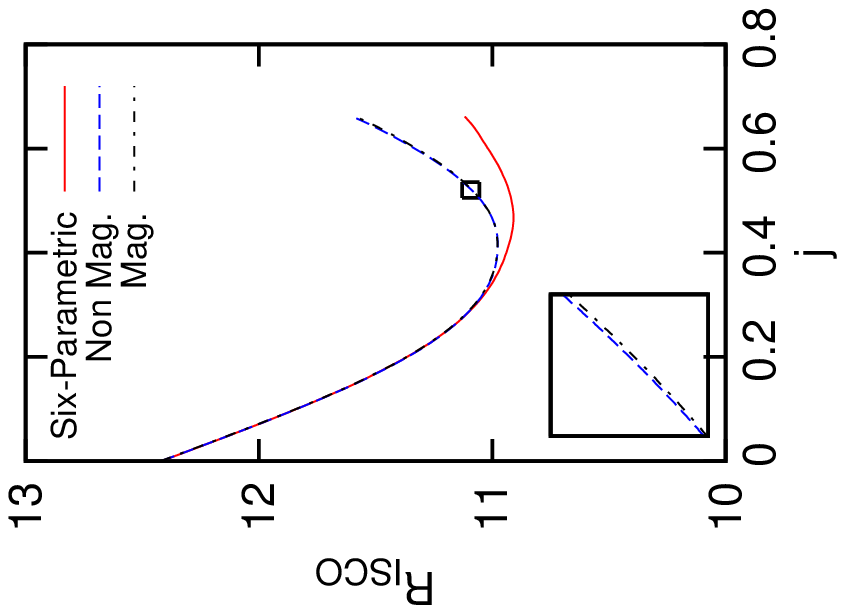}}
&{\includegraphics[angle=270,scale=0.65]{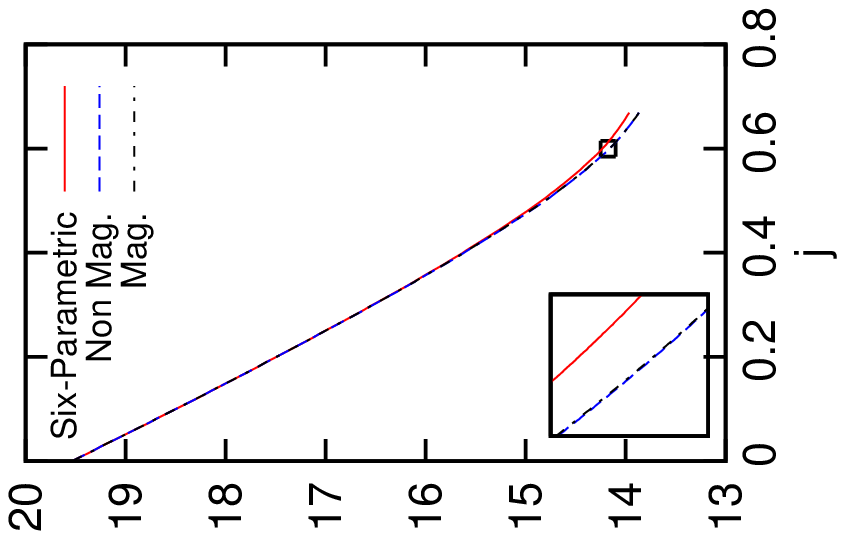}}
\end{tabular}
\end{center}
\caption{Case ${\cal M}_{\rm{nat}} = 1{\rm{km}^2}$: Radius at ISCO (henceforth measured in Km) for the EOS AU sequence with constant rest mass corresponding to a non--rotating model of $1.578 M_{\odot}$ for prograde orbits (panel a) and for EOS APRb sequence with constant rest mass corresponding to a sequence that terminates at the maximum--mass non--rotating model in the
non--rotating limit of $2.672M_{\odot}$ for prograde orbits (panel b). The difference between the Mag. and Non--Mag. cases is negligible, as depicted in the enlargements (insets). Color insets online.}
\label{Fig:mu1}
\end{figure}

For the particular case ${\cal M} = 1{\rm{km}^2} $, in Fig.\ref{Fig:mu1} we plot the radii at ISCO for the six parametric exact solution presented by Pach\'on {\it et al.} (Six-Parametric), the approximate formula (\ref{RISCO}) with magnetic dipole (Mag) and the approximate formula without magnetic dipole (Non-Mag). Here, we use the data of B\&S for EOS AU sequence with constant rest mass corresponding to a non--rotating model of $1.578 M_{\odot}$ (panel a) and the EOS APRb sequence with constant rest mass corresponding to the maximum-mass model in the non--rotating limit $2.672M_{\odot}$  (panel b).  As can be easily noted from the insets in Fig.\ref{Fig:mu1}, for the case of a dipolar magnetic moment of  $10^{31}{\rm{A m}^2}$, the changes introduced in the radius at ISCO by the  approximate formula (\ref{RISCO}) taking into account  the magnetic dipole are negligible.

\begin{figure}[h!]
\begin{center}
\begin{tabular}{cc}
\hspace{0.7cm}(a)&(b)\\
{\includegraphics[angle=270,scale=0.65]{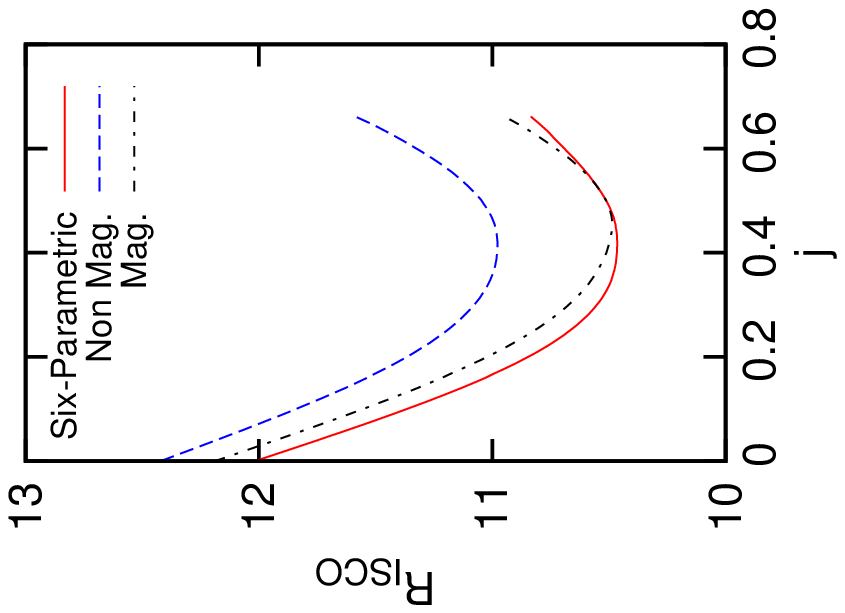}}
&{\includegraphics[angle=270,scale=0.65]{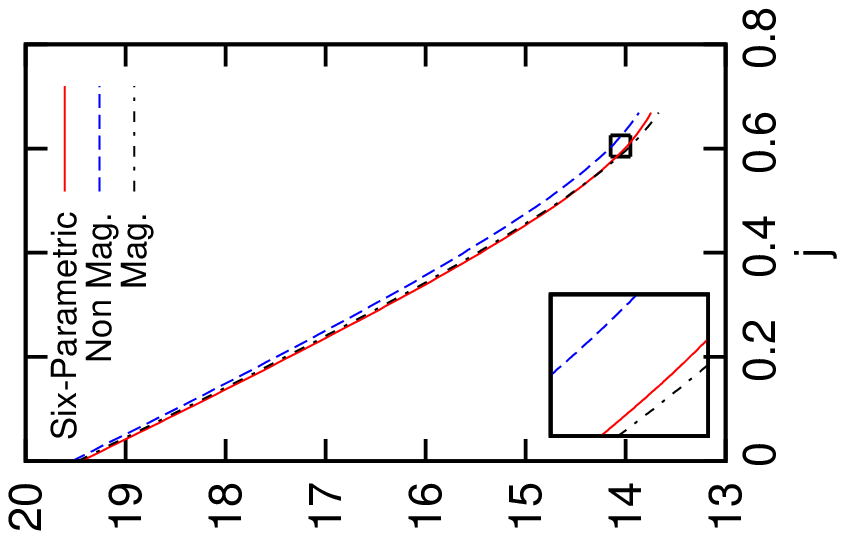}}
\end{tabular}
\end{center}
\caption{Case ${\cal M}= 10{\rm{km}^2}$: Same EOS as in Fig.\ref{Fig:mu1}. The difference between the Mag. and Non--Mag. cases  is appreciably more marked. Color insets online.}
\label{Fig:mu10}
\end{figure}

In Fig.\ref{Fig:mu10} we plot the radii at ISCO for the same parameters as in Fig.\ref{Fig:mu1}, but using the high value ${\cal M}= 10{\rm{km}^2}$. For EOS AU (panel a) the error for the case $q_{\rm{max}} \sim 0.7$ is around $4\%$ for the Non--Mag. case, while it is just close to $1\%$ for Mag. For EOS APRb (panel b) the errors are $1\%$ for Non--Mag. and $0.5\%$ for Mag. Therefore, it can be concluded that for very strong magnetic fields (ten times larger than the ones observed for magnetars) the influence of the magnetic field on the radius at ISCO is significantly important, producing a decreasing tendency on the radius at ISCO with the increasing of the magnetic dipole strength.

\begin{figure}[h!]
\begin{center}
\includegraphics[angle=270,scale=0.6]{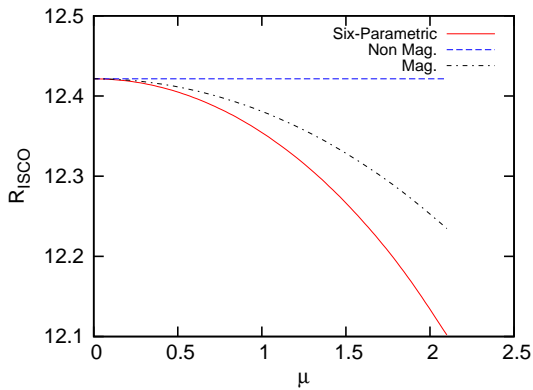}
\caption{Dependence of $R_{\rm ISCO}$ with $\mu$ for parameters of observed magnetars. Note that when $\mu$ becomes larger the radius becomes smaller and the formula with magnetic dipole (labeled as Mag.) is close to the analytical solution (labeled as Six--parametric.) }
\label{Fig:vs.mu}
\end{center}
\end{figure}

Finally, just to illustrate the influence of the magnetic dipole  ${\cal M}$ on the radius at ISCO for magnetars, in Fig.\ref{Fig:vs.mu} we plot the radii at ISCO for the six-parametric exact solution, the magnetized approximate solution and the non-magnetized one. Here we set the multipole moments in accordance with typical values for observed magnetars: $M=1.402 M_{\odot}$, $q={\cal Q}_2=  {\cal Q}_3={\cal Q}_4=10^{-3}$. It can be seen that the radius at ISCO is affected by the existence of magnetic dipole, with an appreciable variation with respect to the non-magnetized case. At $\mu \sim 2 $ (corresponding to a value of the magnetic dipole of about $0.5 \times 10^{32}{\rm{A m}^2}$), the errors are close to 3\% for the non-magnetized version and around 1\% for the magnetized version.  With this example we intent to show that the same tendency as discussed above holds for realistic values of magnetars.

From a mathematical point of view, the strong influence of the magnetic dipole for $\mu\geq 1$, can be explained taking into account that the magnetic dipole parameter $\mu$ appears into the metric functions by means of  $|\Phi^2|$ and products of imaginary and real parts of $\Phi$, so for the dimensionless magnetic dipole parameter $\mu$ its lower order in the metric functions and therefore in the ISCO formulas is of the order $\mu^2$.  From a physical point of view, it is a well known fact that in general relativity the magnetic field could induce observable effects into the space--time (see e.g. \cite{B06} and references therein) and therefore changes onto the dynamics of non magnetized neutral particles orbiting around such sources can be expected \cite{DPS07}.
\section {Concluding Remarks}\label{sec:Conclusions}
In this paper, we have obtained simple approximate formulas for the radius, the angular velocity,
the energy and the angular momentum of a neutral test particle at the innermost stable circular
orbit (ISCO). These formulas correct the results obtained by Shibata and Sasaki \cite{S&S} and generalize
it to the case of neutral test particles moving on the equatorial plane around a rotating source
endowed with magnetic dipole. In order to test the accuracy of our approximate expressions, we first
have compared it with the radius at ISCO (calculated for non-magnetized neutron stars)
by means of the Kerr solution, the six-parametric exact solution given by Pach\'on {\it et al.} \cite{pachon} and the
numerical data given by Berti and Stergioulas  \cite{B&S}  for selected EOS. In all cases, our formula differs
from the numerical results by at most $0.6\%$ in the slow-rotation regime ({\it i.e.} for $j \leq 0.3$). As a main result, it was found that when using realistic parameters for magnetars (including the
magnetic dipole moment of the source), the radius at ISCO exhibits appreciable deviations with a tendency to decrease away from its non-magnetized version, only if the magnetic field strength is higher than $100{\rm GT}$.
Finally, we want to remark that theoretical models for slowly rotating magnetars including numerical data for the higher multipolar structure and the radius at ISCO are desirable.
\section*{ACKNOWLEDGMENTS}
We would like to thank Jes\'us Mart\'in, Eduardo Ruiz  and Leonardo A. Pach\'on for helpful discussions and comments and to the anonymous referee for suggestions and comments that helped us to significantly improve the manuscript. During the research for and writing of this manuscript, J. L. H. P. was partially funded by the Spanish Ministry of Education and Sciences under Research Project No. FIS2006-05319, J. D. S. G.  was supported by the Fundaci\'on Carolina in Spain and F. L. D. was partially funded by Colciencias, DAAD in the program ALECOL and Universidad Nacional de Colombia.

\appendix*
\section*{APPENDIX: COEFFICIENTS FOR THE ERNST POTENTIALS ON THE SYMMETRY AXIS.}
The following expressions show the data on the symmetry axis $m_i:=a_{0,i}$  and $q_i:=b_{0,i}$, obtained by using the procedure of FHP \cite{F&H&P} with the formulas given by Sotiriou and Apostolatos \cite{S&A}, in order to get a source with the following multipole structure: Mass $M_0=M$, angular momentum $M_1= q M^2$, mass quadrupole $M_2=-{\cal Q}_2 M^3$, current octupole moment $M_3=- {\cal Q}_3 M^4$, $2^4$-pole of mass $M_4={\cal Q}_4 M^5$ and magnetic dipole moment  ${\cal M}= \mu M^2$ (all the other multipolar moments are set to zero),
\begin{equation}
a_{0,0}=M, \quad a_{0,1}=\I M^2 q,   \nonumber
\end{equation}
\begin{equation}
a_{0,2}=-M^3 {\cal Q}_2, \quad  a_{0,3}= -\I M^4 {\cal Q}_3, \nonumber
\end{equation}
\begin{equation}
a_{0,4}=\frac{1}{70} M^5 \left(10 q^2-3 \mu ^2-10 {\cal Q}_2+70 {\cal Q}_4 \right), \nonumber
\end{equation}
\begin{equation}
a_{0,5}=-\I \frac{1}{21} M^6 \left(q^3-\mu ^2 q-8 {\cal Q}_2 q+7 {\cal Q}_3\right), \nonumber
\end{equation}
\begin{eqnarray}
a_{0,6}= &&M^7 \left(\frac{23 {\cal Q}_2 q^2}{231}+\frac{q^2}{21}-\frac{14 {\cal Q}_3 q}{33}-\frac{17
{\cal Q}_2^2}{77} \right. \nonumber \\
&& \left. -\frac{205 {\cal Q}_2 \mu ^2}{4158} -\frac{32 \mu^2}{945}-\frac{{\cal Q}_2}{21}+\frac{6 {\cal Q}_4}{11}\right),  \nonumber
\end{eqnarray}
\begin{eqnarray}
a_{0,7}=&& \I M^8 \left(-\frac{13 q^3}{231}+\frac{46 \mu ^2 q}{10395}+\frac{16 {\cal Q}_2 q}{77}-\frac{5 {\cal Q}_3}{33}\right) \nonumber \\
&&+ M^8 {\cal O}(\epsilon^5), \nonumber
\end{eqnarray}
\begin{eqnarray}
a_{0,8}= && M^9 \left( -\frac{40 q^4}{3003}+\frac{2623 \mu ^2 q^2}{216216}+\frac{536 {\cal Q}_2
q^2}{3003}+\frac{5 q^2}{231} \right. \nonumber \\
&&  -\frac{140 {\cal Q}_3 q}{429}+\frac{257 \mu^4}{216216}-\frac{461 {\cal Q}_2^2}{3003}-\frac{137 {\cal Q}_2 \mu ^2}{90090} \nonumber \\
&& \left. -\frac{985 \mu^2}{36036}-\frac{5 {\cal Q}_2}{231}+\frac{45 {\cal Q}_4}{143}\right) + M^9 {\cal O} (\epsilon^5), \nonumber
\end{eqnarray}
\begin{eqnarray}
a_{0,9}=&& \I M^{10} \left(-\frac{115 q^3}{3003}-\frac{713 \mu ^2 q}{135135}+\frac{120 {\cal Q}_2 q}{1001}-\frac{35 {\cal Q}_3}{429}\right) \nonumber \\ && + M^{10}{\cal O} (\epsilon^5), \nonumber
\end{eqnarray}
\begin{eqnarray}
a_{0,10}=&&M^{11} \left(-\frac{115 q^4}{7007}-\frac{26808961 \mu ^2 q^2}{2412159750}+\frac{3049 {\cal Q}_2 q^2}{21021} \right. \nonumber \\
&& +\frac{5 q^2}{429}-\frac{98 {\cal Q}_3 q}{429}+\frac{14777834 \mu^4}{1206079875}-\frac{2018 {\cal Q}_2^2}{21021}\nonumber \\
&& \left. +\frac{633179 {\cal Q}_2 \mu^2}{21441420}-\frac{7132 \mu ^2}{328185}-\frac{5 {\cal Q}_2}{429}+\frac{28{\cal Q}_4}{143}\right) \nonumber \\
&& + M^{11} {\cal O}(\epsilon^5), \nonumber
\end{eqnarray}
\begin{eqnarray}
a_{0,11}= && \I M^{12} \left(-\frac{q^3}{39}-\frac{4914823 \mu ^2 q}{689188500}+\frac{32 {\cal Q}_2q}{429}-\frac{7 {\cal Q}_3}{143}\right) \nonumber \\
&&+ M^{12} {\cal O} (\epsilon^5), \nonumber
\end{eqnarray}
\begin{eqnarray}
a_{0,12}=&&M^{13} \left(-\frac{1569 q^4}{119119}-\frac{27445366301 \mu ^2 q^2}{916620705000}-\frac{392 {\cal Q}_3 q}{2431} \right. \nonumber \\
&&  +\frac{12448{\cal Q}_2 q^2}{119119}+\frac{q^2}{143}+\frac{13085405783 \mu^4}{549972423000}-\frac{38 {\cal Q}_2^2}{637} \nonumber \\
&& +\frac{2754769229 {\cal Q}_2 \mu^2}{54997242300}-\frac{2735479 \mu ^2}{157134978}-\frac{{\cal Q}_2}{143} \nonumber \\
&& \left.+\frac{315{\cal Q}_4}{2431}\right) + M^{13}{\cal O}(\epsilon^5), \nonumber
\end{eqnarray}
\begin{equation}
b_{0,0}=0, \quad b_{0,1}=\I M^2 \mu, \quad b_{0,2}=0, \quad b_{0,3}=0, \nonumber
\end{equation}
\begin{equation}
b_{0,4}=\frac{1}{10} M^5 q \mu, \quad b_{0,5}=\I M^6 \left(\frac{\mu ^3}{21}-\frac{1}{21} q^2 \mu +\frac{5 {\cal Q}_2 \mu }{21}\right), \nonumber
\end{equation}
\begin{equation}
b_{0,6}=M^7 \left(\frac{22 q \mu }{945}+\frac{19 q {\cal Q}_2 \mu }{378}-\frac{{\cal Q}_3 \mu}{6}\right), \nonumber
\end{equation}
\begin{eqnarray}
b_{0,7}=&& \I M^8 \left(-\frac{2 \mu ^3}{297}-\frac{67 q^2 \mu }{1485}-\frac{\mu }{297}+\frac{25 {\cal Q}_2 \mu }{189}\right) \nonumber \\
&& +M^{8} {\cal O}(\epsilon^5), \nonumber
\end{eqnarray}
\begin{eqnarray}
b_{0,8}=&&M^9 \left(-\frac{3137 \mu  q^3}{216216}+\frac{3137 \mu ^3 q}{216216}+\frac{57763 {\cal Q}_2 \mu q}{540540} \right. \nonumber \\
&& \left. +\frac{2749 \mu  q}{540540}-\frac{140 {\cal Q}_3 \mu }{1287}\right) +M^{9} {\cal O}(\epsilon^5),\nonumber
\end{eqnarray}
\begin{eqnarray}
b_{0,9}=&& \I M^{10} \left(-\frac{431 \mu ^3}{30030}-\frac{4124 q^2 \mu }{135135}-\frac{23 \mu}{6435}+\frac{293 {\cal Q}_2 \mu }{3510}\right)\nonumber \\
&& +M^{10} {\cal O}(\epsilon^5), \nonumber
\end{eqnarray}
\begin{eqnarray}
b_{0,10}=&& M^{11} \left(-\frac{3466313 \mu  q^3}{172297125}+\frac{1669477 \mu ^3 q}{344594250} \right. \nonumber \\
&& \left. +\frac{2461259 {\cal Q}_2 \mu  q}{26507250}-\frac{709967 \mu q}{689188500}-\frac{3929 {\cal Q}_3 \mu }{64350}\right) \nonumber \\
&& +M^{11} {\cal O}(\epsilon^5), \nonumber
\end{eqnarray}
\begin{eqnarray}
b_{0,11}=&&\I M^{12} \left(-\frac{334865 \mu ^3}{26189163}-\frac{32283217 q^2 \mu}{1454953500}\right. \nonumber \\
&& \left. -\frac{97613 \mu }{31177575} +\frac{25724987 {\cal Q}_2 \mu}{436486050}\right)+M^{12} {\cal O}(\epsilon^5),\nonumber
\end{eqnarray}
\begin{eqnarray}
b_{0,12}=&&M^{13} \left(-\frac{11686577027 \mu  q^3}{549972423000}+\frac{3542515517 \mu ^3 q}{2749862115000} \right. \nonumber \\ && +\frac{1582913357 {\cal Q}_2 \mu  q}{21152785500}-\frac{24121271 \mu q}{7856748900} \nonumber \\
&& \left . -\frac{20220848 {\cal Q}_3 \mu }{654729075}\right)+M^{13} {\cal O}(\epsilon^5).
\end{eqnarray}


\suppressfloats


\begin{thebibliography}{20}
\bibitem{note}  In the S\&S work \cite{S&S}, they have used the same definition for $M_{40}:=m_4 m_0-m_3 m_1$ as Fodor {\it et al.} \cite{F&H&P} but with a conceptual mistake in their calculation. They have confused the multipole on the symmetry axis $m_4:=a_{0,4}$ with the general  multipole moment $M_4$, as can be seen in the  formula $M_{40}=M_4 M + S_3 J$, below the formula (A1) in Ref. \cite{S&S}.
\end{thebibliography}
\end{document}